\documentclass[runningheads]{llncs}
\usepackage{graphicx}
\usepackage[misc]{ifsym}
\usepackage{url}

\usepackage[utf8]{inputenc}
\usepackage[english]{babel}
\usepackage{bbding}
\usepackage{hyperref}
\usepackage{rotating}
\usepackage{verbatim}
\usepackage{amssymb}
\usepackage{amsmath}
\usepackage{xcolor}
\usepackage{mathtools, nccmath}
\definecolor{commentgreen}{RGB}{2,112,10}
\definecolor{eminence}{RGB}{108,48,130}
\definecolor{weborange}{RGB}{255,165,0}
\definecolor{frenchplum}{RGB}{129,20,83}
\definecolor{sparqlblue}{RGB}{84,153,199}
\usepackage{listings}
\usepackage{multicol}

\usepackage{amsthm}
\theoremstyle{definition}
\usepackage[flushleft]{threeparttable}
\usepackage[font=small,labelfont=bf]{caption}
\usepackage[normalem]{ulem}
\begin{document}
\title{An Ontology for the Materials Design Domain}

\titlerunning{Materials Design Ontology}
%


\author{Huanyu Li \inst{1,3} \orcidID{0000-0003-1881-3969} \and
Rickard Armiento \inst{2,3} \orcidID{0000-0002-5571-0814} \and
Patrick Lambrix \inst{1,3} \Letter \orcidID{0000-0002-9084-0470}}

\institute{
Department of Computer and Information Science, \\
Link{\"o}ping University, 581 83 Link{\"o}ping, Sweden \\
\and
Department of Physics, Chemistry and Biology, \\
Link{\"o}ping University, 581 83 Link{\"o}ping, Sweden \\
\and
The Swedish e-Science Research Centre, Link{\"o}ping University,\\
581 83 Link{\"o}ping, Sweden \\
\email{firstname.lastname@liu.se} 
}

\titlerunning{An Ontology for the Materials Design Domain}
\maketitle 
\setcounter{footnote}{0}
%
\begin{abstract}

In the materials design domain, much of the data from materials calculations are stored in different heterogeneous databases.
Materials databases usually have different data models. Therefore, the users have to face the challenges to find the data from adequate sources and integrate data from multiple sources.
Ontologies and ontology-based techniques can address such problems as the formal representation of domain knowledge can make data more available and interoperable among different systems. In this paper, we introduce the Materials Design Ontology (MDO), which defines concepts and relations to cover knowledge in the field of materials design. MDO is designed using domain knowledge in materials science (especially in solid-state physics), and is guided by the data from several databases in the materials design field. 
We show the application of the MDO to materials data retrieved from well-known materials databases. 

\vspace{5mm} 

\noindent {\bf Resource Type:} Ontology

\noindent {\bf IRI:} \url{https://w3id.org/mdo/full/1.0/}
\end{abstract}

\section{Introduction}
More and more researchers in the field of materials science have realized that data-driven techniques have the potential to accelerate the discovery and design of new materials. Therefore, a large number of research groups and communities have developed data-driven workflows, including data repositories (for an overview see \cite{lambrix2018big}) and task-specific analytical tools.
Materials design is a technological process with many applications. The goal is often to achieve a set of desired materials properties for an application under certain limitations in e.g., avoiding or eliminating toxic or critical raw materials.
The development of condensed matter theory and materials modeling, has made it possible to achieve quantum mechanics-based simulations that can generate reliable materials data by using computer programs \cite{lejaeghere2016reproducibility}. 
For instance, in \cite{armiento2019database} a flow of databases-driven high-throughput materials design in which the database is used to find materials with desirable properties, is shown.
A global effort, the Materials Genome Initiative\footnote{\url{https://www.mgi.gov/}}, has been proposed to govern databases that contain both experimentally-known and computationally-predicted material properties. The basic idea of this effort is that searching materials databases with desired combinations of properties could help to address some of the challenges of materials design. 
As these databases are heterogeneous in nature, there are a number of challenges to using them in the materials design workflow. For instance, retrieving data from more than one database means that users have to understand and use different application programming interfaces (APIs) or even different data models to reach an agreement.
Nowadays, materials design interoperability is achieved mainly via file-based exchange involving specific formats and, at best, some partial metadata, which is not always adequately documented as it is not guided by an ontology.
The second author is closely involved with another ongoing effort, the Open Databases Integration for Materials Design (OPTIMADE\footnote{\url{https://www.optimade.org/}})  project which aims at making materials databases interoperational by developing a common API.
Also this effort would benefit from semantically enabling the system using an ontology, both for search as well as for integrating information from the underlying databases.

These issues relate to the FAIR principles (Findable, Accessible, Interoperable, and Reusable), with the purpose of enabling machines to automatically find and use the data, and individuals to easily reuse the data \cite{wilkinson2016fair}. Also in the materials science domain, recently,  an awareness  regarding the importance of such principles for data storage and management is developing and research in this area is starting \cite{draxl2018nomad}.

To address these challenges and make data FAIR, ontologies and ontology-based techniques have been proposed to play a significant role. 
For the materials design field there is, therefore, a need for an ontology to represent solid-state physics concepts such as  materials' properties,  microscopic structure as well as calculations, which are the basis for materials design. 
Thus, in this paper, we present the Materials Design Ontology (MDO). The development of MDO was guided by the schemas of OPTIMADE as they are based on a consensus reached by several of the materials database providers in the field. Further, we show the use of MDO for data obtained via the OPTIMADE API and via database-specific APIs  in the materials design field. 

The paper is organized as follows. We introduce some well-known databases and existing ontologies in the materials science domain in Section \ref{sec-related-work}. In Section \ref{sec-MDO} we present the development of MDO and introduce the concepts, relations and the  axiomatization of the ontology. In Section \ref{sec-MDO-use} we introduce the envisioned usage  of MDO as well as a current implementation. In Section \ref{sec-discussion} we discuss such things as the impact, availability and extendability  of MDO as well as future work. Finally, the paper concludes in Section \ref{sec-conclusion} with a small summary.

\vspace*{0.2cm}

{\bf Availability:} MDO is developed and maintained on a GitHub repository\footnote{\url{https://github.com/huanyu-li/Materials-Design-Ontology}}, and is available from a permanent w3id URL\footnote{\url{https://w3id.org/mdo}}.

\section{Related Work}
\label{sec-related-work}

In this section we discuss briefly well-known databases as well as ontologies in the materials science field. Further, we briefly introduce OPTIMADE. 

\subsection{Data and Databases in the Materials Design Domain}
\label{sect-db}
In the search for designing new materials, the calculation of electronic structures is an important tool. Calculations take data representing the structure and property of materials as input and generate new such data. A common crystallographic data representation that is widely used by researchers and software vendors for materials design, is CIF\footnote{Crystallographic Information Framework, \url{https://www.iucr.org/resources/cif}}. It was developed by  the International Union of Crystallography Working Party on Crystallographic Information and was first online in 2006. 
One of the widely used databases is the Inorganic Crystal Structure Database (ICSD)\footnote{\url{https://icsd.products.fiz-karlsruhe.de/}}. ICSD provides data that is used as an important starting point in many calculations in the materials design domain. 

As the size of computed data grows, and more and more  machine learning and data mining techniques are being used in materials design, frameworks are appearing that not only provide data but also tools.
Materials Project, AFLOW and OQMD are well-known examples of such frameworks that are publicly available.
\textbf{Materials Project} \cite{Jain2013} is a central program of the Materials Genome Initiative, focusing on predicting the properties of all known inorganic materials through computations. It provides open web-based data access to computed information on materials, as well as tools to design new materials. To make the data publicly available, the Materials Project provides open Materials API and an open-source python-based programming package (pymatgen).
\textbf{AFLOW} \cite{curtarolo2012aflow} (Automatic Flow for Materials Discovery) is an automatic framework for high-throughput materials discovery, especially for crystal structure properties of alloys, intermetallics, and inorganic compounds.  AFLOW provides a REST API and a python-based programming package (aflow).  
\textbf{OQMD}  \cite{saal2013materials} (The Open Quantum Materials Database) is also a high-throughput database consisting of over 600,000 crystal structures calculated based on density functional theory\footnote{\url{http://oqmd.org}}. OQMD is designed based on a relational data model. OQMD supports a REST API and a python-based programming package (qmpy).

\subsection{Ontologies and Standards}
Within the materials science domain, the use of semantic technologies is in its infancy with the development of ontologies and standards. The ontologies have been developed, focusing on representing general materials domain knowledge and specific sub-domains respectively. 

Two ontologies representing general materials domain knowledge and to which our ontology connects are ChEBI and EMMO. \textbf{ChEBI} \cite{degtyarenko2007chebi} (Chemical Entities of Biological Interest) is a freely available data set of molecular entities focusing on chemical compounds. The representation of such molecular entities as atom, molecule ion, etc. is the basis in both chemistry and physics. The ChEBI ontology is widely used and integrated into other domain ontologies.
\textbf{EMMO} (European Materials \& Modelling Ontology) is an upper ontology that is currently being developed and aims at developing a standard representational ontology framework based on current knowledge of materials modeling and characterization.
The EMMO development started from the very bottom level, using the actual picture of the physical world coming from applied sciences, and in particular from physics and material sciences. 
Although EMMO already covers some sub-domains in materials science, many sub-domains are still lacking, including the domain MDO targets. 

Further, a number of ontologies from the materials science domain focus on specific sub-domains (e.g., metals, ceramics, thermal properties, nanotechnology), and have been developed with a specific use in mind (e.g., search, data integration) \cite{lambrix2018big}. For instance, the Materials Ontology \cite{ashino2010materials} was developed for data exchange among thermal property databases, and Mat\-Onto ontology \cite{cheung2008towards} for oxygen ion conducting materials in the fuel cell domain. NanoParticle Ontology \cite{thomas2011nanoparticle} represents properties of nanoparticles with the purpose of designing new nanoparticles, while the eNanoMapper ontology \cite{hastings2015enanomapper} focuses on assessing risks related to the use of nanomaterials from the engineering point of view. Extensions to these ontologies in the nanoparticle domain are presented in \cite{li2019method}.
An ontology that represents formal knowledge for simulation, modeling, and optimization in computational molecular engineering is presented in \cite{horsch2019semantic}. Further, an ontology design pattern to model material transformation in the field of sustainable construction, is proposed in \cite{VardemanKCJFHB17}. All the materials science domain ontologies above target different sub-domains from MDO.

There are also efforts on building standards for data export from databases and data integration among tools. To some extent the standards formalize the description of materials knowledge and thereby create ontological knowledge.
A recent approach is Novel Materials Discovery (NOMAD\footnote{\url{https://www.nomad-coe.eu/externals/european-centres-of-excellence}}) \cite{draxl2019nomad} of which the metadata structure is defined to be independent of specific material science theory or methods that could be used as an exchange format \cite{Ghiringhelli16}.

\subsection{Open Databases Integration for Materials Design}
OPTIMADE is a consortium gathering many database providers. It aims at enabling interoperability between materials databases through a common REST API. During the development OPTIMADE takes widely used materials databases such as those introduced in section \ref{sect-db} into account. 
OPTIMADE has a schema that defines the specification of the OPTIMADE REST API and provides essentially a list of terms for which there is a consensus from different database providers. The OPTIMADE API is taken into account in the development of MDO as shown in section \ref{sec-MDO}.

\section{The Materials Design Ontology (MDO)}
\label{sec-MDO}

\subsection{The development of MDO}
The development of MDO followed the NeOn ontology engineering methodology \cite{suarez2012neon}. It consists of a number of scenarios mapped from a set of common ontology development activities. In particular, we focused on applying scenario 1 (\textit{From Specification to Implementation}), scenario 2 (\textit{Reusing and re-engineering non-ontological resources}), scenario 3 (\textit{Reusing ontological resources}) and scenario 8 (\textit{Restructuring ontological resources}).
We used OWL2 DL as the representation language for MDO.
During the whole process, two knowledge engineers, and one domain expert from the materials design domain were involved. 
In the remainder of this section, we introduce the key aspects of the development of MDO.

\subsubsection{Requirements Analysis.} During this step, we clarified the requirements by proposing Use Cases (UC), Competency Questions (CQ) and additional restrictions. 

The use cases, which were identified through literature study and discussion between the domain expert and the knowledge engineers based on experience with the development of OPTIMADE and the use of materials science databases, are listed below. 
\begin{itemize}
    \item UC1: MDO will be used for representing knowledge in basic materials science such as solid-state physics and condensed matter theory.
    \item UC2: MDO will be used for representing materials calculation and standardizing the publication of the materials calculation data.
    \item UC3: MDO will be used as a standard to improve the interoperability among heterogeneous databases in the materials design domain.
    \item UC4: MDO will be mapped to OPTIMADE's schema to improve OPTIMADE's search functionality. 
\end{itemize}

The competency questions are based on discussions with domain experts and contain questions that the databases currently can answer as well as questions that experts would want to ask the databases. For instance, CQ1, CQ2, CQ6, CQ7, CQ8 and CQ9 cannot be asked explicitly through the database APIs, although the original downloadable data contains the answers.
\begin{itemize}
    \item CQ1: What are the calculated properties and their values produced by a materials calculation?
    \item CQ2: What are the input and output structures of a materials calculation?
    \item CQ3: What is the space group type of a structure?
    \item CQ4: What is the lattice type of a structure?
    \item CQ5: What is the chemical formula of a structure?
    \item CQ6: For a series of materials calculations, what are the compositions of materials with a specific range of a calculated property (e.g., band gap)?
    \item CQ7: For a specific material and a given range of a calculated property (e.g., band gap), what is the lattice type of the structure?
    \item CQ8: For a specific material and an expected lattice type of output structure, what are the values of calculated properties of the calculations?
    \item CQ9: What is the computational method used in a materials calculation?
    \item CQ10: What is the value for a specific parameter (e.g., cutoff energy) of the method used for the calculation?
    \item CQ11: Which software produced the result of a calculation?
    \item CQ12: Who are the authors of the calculation?
    \item CQ13: Which software or code does the calculation run with?
    \item CQ14: When was the calculation data published to the database?
\end{itemize}

Further, we proposed a list of additional restrictions that help in defining concepts. Some examples are shown below.
The full list of additional restrictions can be found at the GitHub repository\footnote{\url{https://github.com/huanyu-li/Materials-Design-Ontology/blob/master/requirements.md}}.
\begin{itemize}
    \item A materials property can relate to a structure.
    \item A materials calculation has exactly one corresponding computational method.
    \item A structure corresponds to one specific space group.
    \item A materials calculation is performed by some software programs or codes.
\end{itemize}

\subsubsection{Reusing and re-engineering non-ontological resources.}
To obtain the knowledge for building the ontology, we followed two steps: (1) the collection and analysis of non-ontological resources that are relevant to the materials design domain, and (2) discussions with the domain expert regarding the concepts and relationships to be modeled in the ontology. The collection of non-ontological resources comes from: (1) the dictionaries of CIF and International Tables for Crystallography; (2) the APIs from different databases (e.g., Materials Project, AFLOW, OQMD) and OPTIMADE.

\subsubsection{Modular development aiming at building design patterns.} We identified a pattern related to provenance information in the repository of Ontology Design Patterns (ODPs) that could be reused or re-engineered for MDO. This has led to the reuse of entities in PROV-O \cite{lebo2013prov1}.
Further, we built MDO in modules considering the possibility for each module to be an ontology design pattern, e.g., the calculation module. 

\subsubsection{Connection and Integration of Existing Ontologies.}
MDO is connected to EMMO by reusing the concept `Material', and to ChEBI by reusing the concept `atom'. Further, we reuse the concepts `Agent' and `SoftwareAgent' from PROV-O.
In terms of representation of units we reuse the `Quantity', `QuantityValue', `QuantityKind' and `Unit' concepts from QUDT (Quantities, Units, Dimensions and Data Types Ontologies) \cite{haasquantities}.
We use the metadata terms from the Dublin Core Metadata Initiative (DCMI)\footnote{\url{http://purl.org/dc/terms/}} to represent the metadata of MDO. 

\subsection{Description of MDO}

\begin{figure}[hbt!] 
\centering
\includegraphics[width=1.0\textwidth]{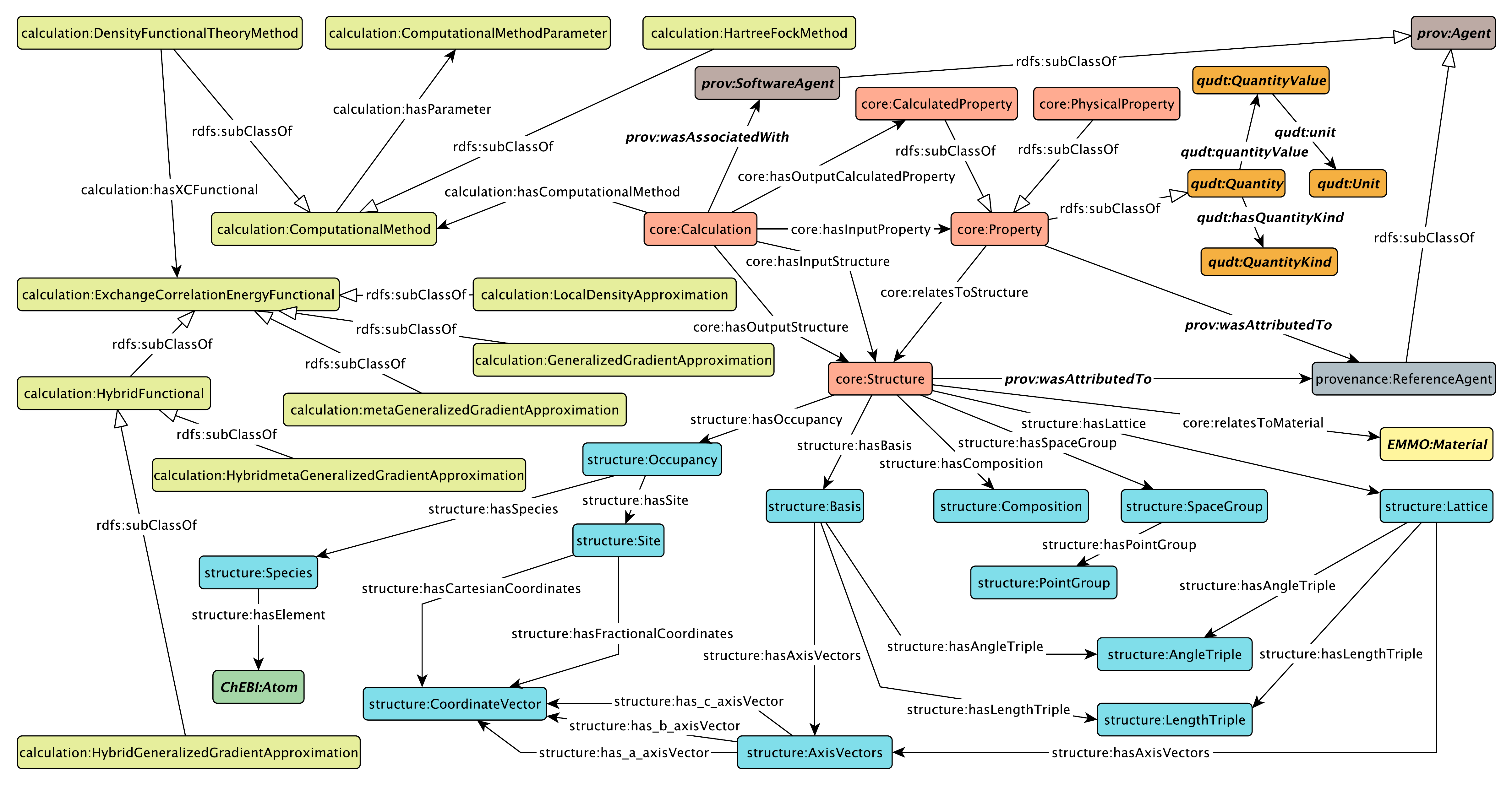}
\caption{An overview of MDO.}
\label{fig:mdo}
\end{figure}

MDO consists of one basic module, \textit{Core}, and two domain-specific modules, \textit{Structure} and \textit{Calculation},  importing the \textit{Core} module. In addition, the \textit{Provenance} module, which also imports \textit{Core}, models provenance information. 
In total, the OWL2 DL representation of the ontology contains 37 classes, 32 object properties, and 32 data properties. Figure \ref{fig:mdo} shows an overview of the ontology. 
The ontology specification is also publicly accessible at w3id.org\footnote{ \url{https://w3id.org/mdo/full/1.0/}}.
The competency questions can be answered using the concepts and relations in the different modules (CQ1 and CQ2 by \textit{Core}, CQ3 to CQ8 by \textit{Structure}, CQ9 and CQ10 by \textit{Calculation}, and CQ11 to CQ14 by \textit{Provenance}).

\begin{figure}[hbt!] 
\centering
\includegraphics[width=1.0\textwidth]{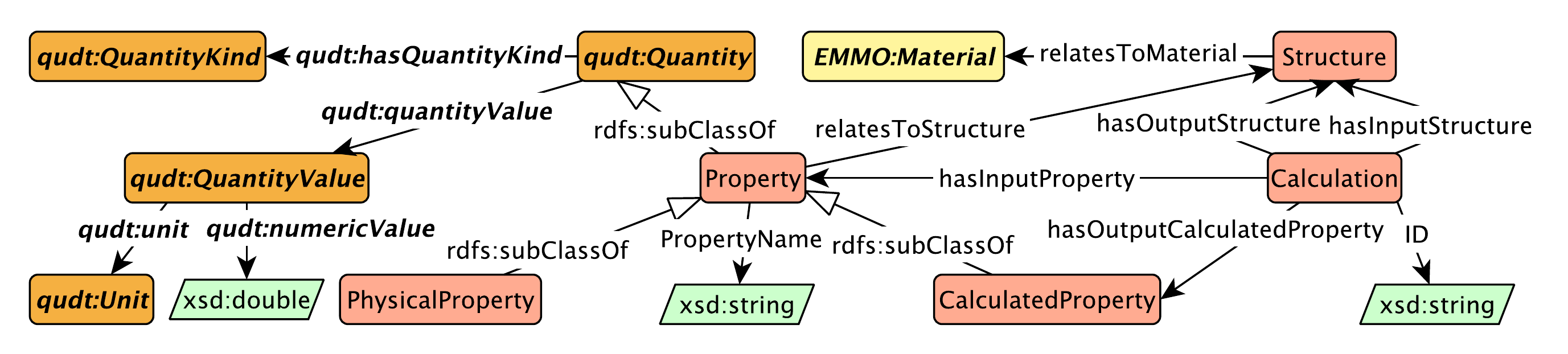}
\caption{Concepts and relations in the Core module.}
\label{fig:core-module}
\end{figure}

The {\bf Core} module as shown in Figure \ref{fig:core-module}, consists of the top-level concepts and relations of MDO, which are also reused in other modules. Figure \ref{fig:core-axioms} shows the description logic axioms for the \textit{Core} module.
The module represents  general information of materials calculations. 
The concepts \textit{Calculation} and \textit{Structure} represent materials calculations and materials' structures, respectively, while \textit{Property} represents materials properties.
\textit{Property} is specialized into the disjoint concepts \textit{CalculatedProperty} and \textit{PhysicalProperty} (Core1, Core2, Core3). 
\textit{Property}, which can be viewed as a quantifiable aspect of one material or materials system,  is defined as a sub concept of \textit{Quantity} from QUDT (Core4).
 \textit{Properties} are also related to \textit{structures} (Core5).
When a calculation is applied on materials structures, each \textit{calculation} takes some \textit{structures} and \textit{properties} as input, and may output \textit{structures} and \textit{calculated properties} (Core6, Core7). 
Further, we use EMMO's concept \textit{Material} and state that each \textit{structure} is related to some \textit{material} (Core8). 

\begin{figure}[hbt!] 
\begin{small} 
\begin{it}
\noindent 
(Core1)   CalculatedProperty $\sqsubseteq$Property \\ 
(Core2) PhysicalProperty $\sqsubseteq$ Property\\
(Core3)     CalculatedProperty $\sqcap$ PhysicalProperty $\sqsubseteq$ $\perp$ \\
(Core4)    Property $\sqsubseteq$ Quantity \\
(Core5)    Property $\sqsubseteq$ $\forall$ relatesToStructure.Structure  \\
(Core6)    Calculation $\sqsubseteq$ $\exists$ hasInputStructure.Structure 
    $\sqcap$ $\forall$   hasInputStructure.Structure\\
\hspace*{1.2cm}    $\sqcap$ $\forall$   hasOutputStructure.Structure \\
(Core7)    Calculation $\sqsubseteq$ $\exists$ hasInputProperty.Property 
    $\sqcap$ $\forall$  hasInputProperty.Property \\
\hspace*{1.2cm}      $\sqcap$ $\forall$  hasOutputCalculatedProperty.CalculatedProperty \\   
(Core8)    Structure $\sqsubseteq$ $\exists$ relatesToMaterial.Material $\sqcap$ $\forall$ relatesToMaterial.Material\\
\end{it}
\end{small}
\caption{Description logic axioms for the Core module.}
\label{fig:core-axioms}
\end{figure}

\begin{figure}[hbt!] 
\centering
\includegraphics[width=1.0\textwidth]{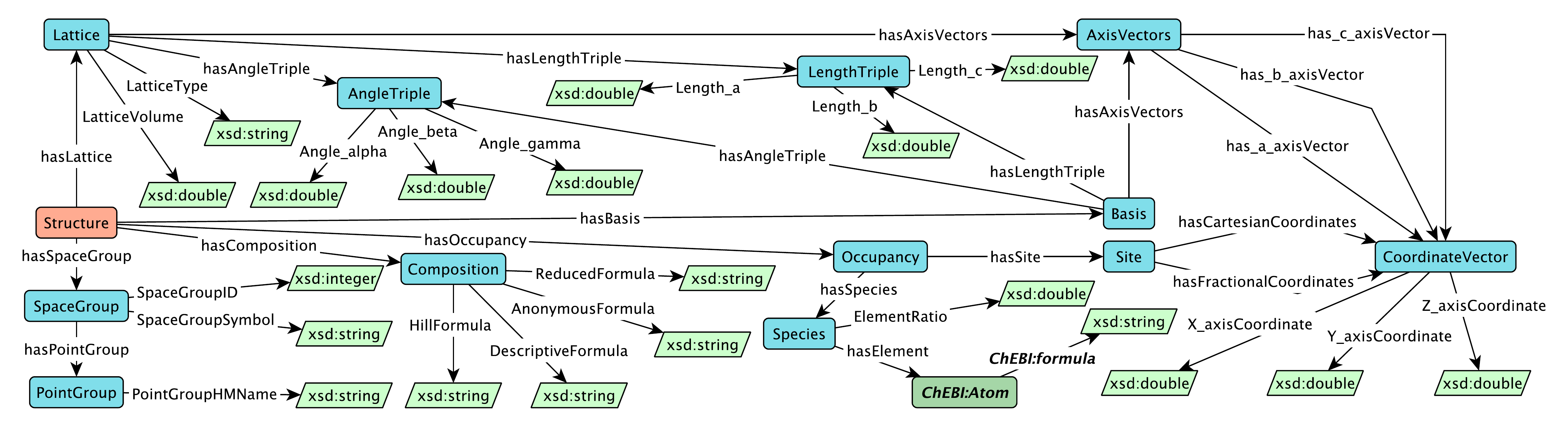}
\caption{Concepts and relations in the Structure module.}
\label{fig:structure-module}
\end{figure}

The {\bf Structure} module as shown in Figure \ref{fig:structure-module}, represents the structural information of materials. 
Figure \ref{fig:structure-axioms} shows the description logic axioms for the \textit{Structure} module.
Each \textit{structure} has exact one \textit{composition} which represents what chemical elements compose the structure and the ratio of elements in the \textit{structure} (Struc1). The \textit{composition} has different representations of chemical formulas. The \textit{occupancy} of a structure relates the \textit{sites} with the \textit{species}, i.e. the specific chemical elements, that occupy the \textit{site} (Struc2 - Struc5).
Each \textit{site} has at most one representation of coordinates in Cartesian format and at most one in fractional format (Struc6, Struc7).
The spatial information regarding structures is essential to reflect  physical characteristics such as melting point and strength of materials.
To represent this spatial information, we state that each \textit{structure} is represented by some \textit{bases} and a (periodic) \textit{structure}  can also be represented by one or more \textit{lattices} (Struc8). 
Each \textit{basis} and each \textit{lattice}  can be identified by one \textit{axis-vectors} set or one \textit{length triple} together with one \textit{angle triple} (Struc9, Struc10). 
An \textit{axis-vectors} set has three connections to \textit{coordinate vector} representing the coordinates of three translation vectors respectively, which are used to represent a (minimal) repeating unit (Struc11). These three translation vectors are often called  a, b, and c.
Point groups and space groups are used to represent information of the symmetry of a structure. The \textit{space group} represents a symmetry group of patterns in three dimensions of a \textit{structure} and the \textit{point group} represents a group of linear mappings which correspond to the group of motions in space to determine the symmetry of a \textit{structure}.
Each \textit{structure} has one corresponding \textit{space group} (Struc12).  Based on the definition from International Tables for Crystallography, each \textit{space group} also has some corresponding \textit{point groups} (Struc13). 

\begin{figure}[hbt!] 
\begin{small} 
\begin{it}
\noindent 
(Struc1) Structure $\sqsubseteq$ = 1 hasComposition.Composition \\
\hspace*{1.2cm}  $\sqcap$ $\forall$ hasComposition.Composition\\
(Struc2) Structure $\sqsubseteq$ $\exists$ hasOccupancy.Occupancy $\sqcap$ $\forall$ hasOccupancy.Occupancy \\
(Struc3) Occupancy $\sqsubseteq$ $\exists$ hasSpecies.Species $\sqcap$ $\forall$ hasSpecies.Species \\
(Struc4) Occupancy $\sqsubseteq$ $\exists$ hasSite.Site $\sqcap$ $\forall$ hasSite.Site\\
(Struc5) Species $\sqsubseteq$ = 1 hasElement.Atom \\
(Struc6) Site $\sqsubseteq$ $\leq$ 1 hasCartesianCoordinates.CoordinateVector \\
\hspace*{1.2cm}  $\sqcap$ $\forall$ hasCartesianCoordinates.CoordinateVector\\
(Struc7)   Site $\sqsubseteq$ $\leq$ 1  hasFractionalCoordinates.CoordinateVector \\
\hspace*{1.2cm}  $\sqcap$ $\forall$ hasFractionalCoordinates.CoordinateVector \\
(Struc8) Structure $\sqsubseteq$ $\exists$ hasBasis.Basis $\sqcap$ $\forall$ hasBasis.Basis $\sqcap$ $\forall$ hasLattice.Lattice\\
(Struc9) Basis $\sqsubseteq$ = 1 hasAxisVectors.AxisVectors $\sqcup$ \\  
\hspace*{1.2cm}  (= 1 hasLengthTriple.LengthTriple $\sqcap$  = 1 hasAngleTriple.AngleTriple) \\
(Struc10) Lattice $\sqsubseteq$ = 1 hasAxisVectors.AxisVectors $\sqcup$  \\ 
\hspace*{1.2cm}  (= 1 hasLengthTriple.LengthTriple $\sqcap$  = 1 hasAngleTriple.AngleTriple) \\
(Struc11) AxisVectors $\sqsubseteq$ = 1 has\_a\_axisVector.CoordinateVector \\
\hspace*{1.2cm} $\sqcap$ = 1 has\_b\_axisVector.CoordinateVector \\
\hspace*{1.2cm} $\sqcap$ = 1 has\_c\_axisVector.CoordinateVector \\
(Struc12) Structure $\sqsubseteq$ = 1 hasSpaceGroup.SpaceGroup $\sqcap$ $\forall$ hasSpaceGroup.SpaceGroup \\
(Struc13) SpaceGroup $\sqsubseteq$ $\exists$ hasPointGroup.PointGroup $\sqcap$ $\forall$ hasPointGroup.PointGroup \\
\end{it}
\end{small}
\caption{Description logic axioms for the Structure module.}
\label{fig:structure-axioms}
\end{figure}

\begin{figure}[hbt!] 
\centering
\includegraphics[width=1.0\textwidth]{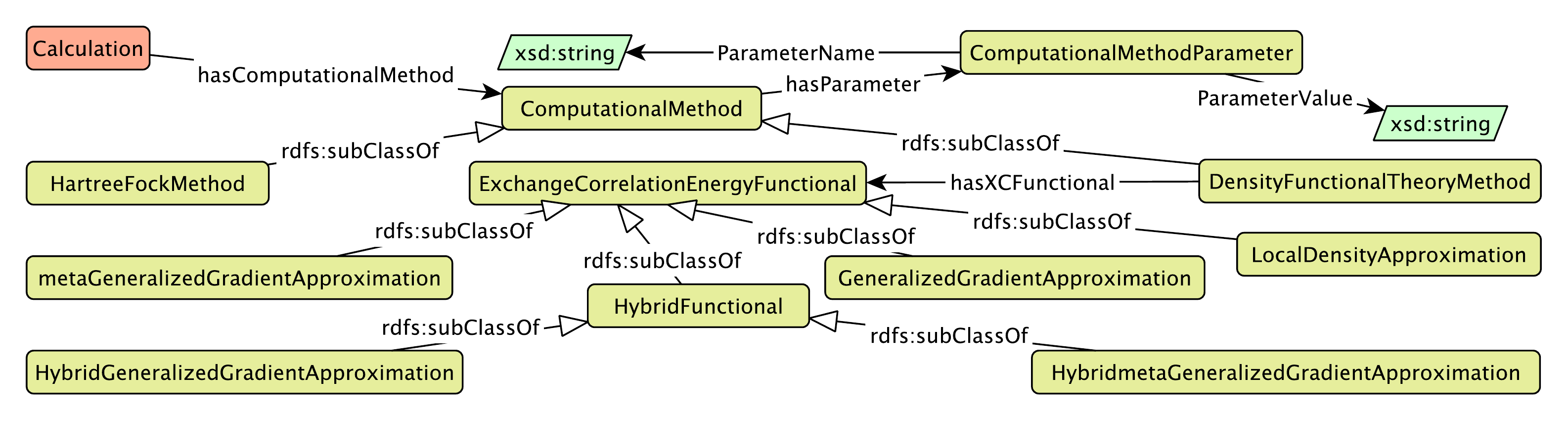}
\caption{Concepts and relations in the Calculation module.}
\label{fig:calculation-module}
\end{figure}

The {\bf Calculation} module as shown in Figure \ref{fig:calculation-module},
represents the classification of different computational methods. 
Figure \ref{fig:calculation-axioms} shows the description logic axioms for the \textit{Calculation} module.
Each \textit{calculation} is achieved by a specific \textit{computational method} (Cal1). 
Each \textit{computational method} has some \textit{parameters} (Cal2). In the current version of this module, we represent two different methods, the  \textit{density functional theory method} and the \textit{HartreeFock method} (Cal3, Cal4). 
In particular, the density functional theory method is frequently used in materials design to investigate the electronic structure. 
Such method has at least one corresponding \textit{exchange correlation energy functional} (Cal5) which is used to calculate the exchange-correlation energy of a system.
There are different kinds of functionals to calculate exchange–correlation energy (Cal6 - Cal11). 

\begin{figure}[hbt!] 
\begin{small} 
\begin{it}
\noindent 
(Cal1)         Calculation $\sqsubseteq$ = 1 hasComputationalMethod.ComputationalMethod \\
(Cal2) ComputationalMethod $\sqsubseteq$ $\exists$ hasParameter.ComputationalMethodParameter \\ 
\hspace*{1.0cm}  $\sqcap$ $\forall$ hasParameter.ComputationalMethodParameter\\
(Cal3)     DensityFunctionalTheoryMethod $\sqsubseteq$ ComputationalMethod\\ 
(Cal4) HartreeFockMethod $\sqsubseteq$ ComputationalMethod  \\
(Cal5)     DensityFunctionalTheoryMethod $\sqsubseteq$ \\
\hspace*{1.0cm} $\exists$ hasXCFunctional.ExchangeCorrelationEnergyFunctional \\
\hspace*{1.0cm}  $\sqcap$ $\forall$ hasXCFunctional.ExchangeCorrelationEnergyFunctional \\
(Cal6)     GeneralizedGradientApproximation $\sqsubseteq$ ExchangeCorrelationEnergyFunctional \\ 
 (Cal7) LocalDensityApproximation $\sqsubseteq$ ExchangeCorrelationEnergyFunctional \\
(Cal8)    metaGeneralizedGradientApproximation $\sqsubseteq$ \\
\hspace*{1.0cm} ExchangeCorrelationEnergyFunctional \\
(Cal9) HybridFunctional $\sqsubseteq$ ExchangeCorrelationEnergyFunctional \\
(Cal10) HybridGeneralizedGradientApproximation $\sqsubseteq$
HybridFunctional \\
(Cal11) HybridmetaGeneralizedGradientApproximation $\sqsubseteq$
HybridFunctional \\
\end{it}
\end{small}
\caption{Description logic axioms for the Calculation module.}
\label{fig:calculation-axioms}
\end{figure}

\begin{figure}[hbt!] 
\centering
\includegraphics[width=1.0\textwidth]{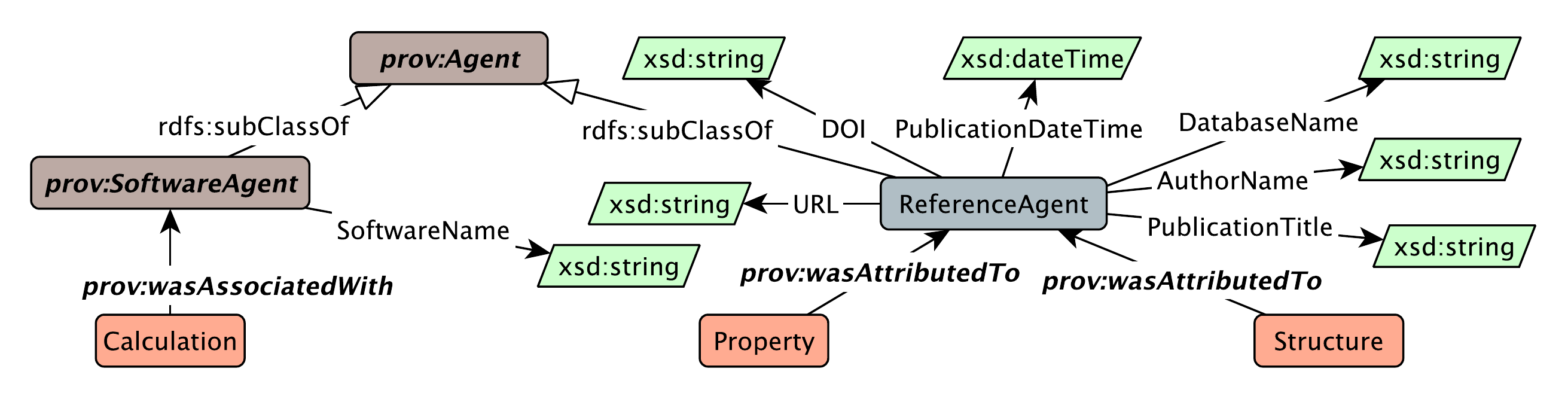}
\caption{Concepts and relations in the Provenance module.}
\label{fig:provenance-module}
\end{figure}

The {\bf Provenance} module as shown in Figure \ref{fig:provenance-module},
represents the provenance information of materials data and calculation. 
Figure \ref{fig:provenance-axioms} shows the description logic axioms for the \textit{Provenance} module.
We reuse part of PROV-O and define a new concept \textit{ReferenceAgent} as a sub-concept of PROV-O's agent (Prov1). 
We state that each \textit{structure} and \textit{property} can be published by \textit{reference agents} which could be databases or  publications (Prov2, Prov3).
Each \textit{calculation} is produced by a specific \textit{software} (Prov4).

\begin{figure}[hbt!]
\begin{small} 
\begin{it}
\noindent 
(Prov1) ReferenceAgent $\sqsubseteq$ Agent \\
(Prov2)     Structure $\sqsubseteq$  $\forall$ wasAttributedTo.ReferenceAgent \\
(Prov3)     Property $\sqsubseteq$  $\forall$ wasAttributedTo.ReferenceAgent \\
(Prov4)    Calculation $\sqsubseteq$  $\exists$ wasAssociatedwith.SoftwareAgent \\
\end{it}
\end{small}
\caption{Description logic axioms for the Provenance module.}
\label{fig:provenance-axioms}
\end{figure}

\section{MDO Usage}
\label{sec-MDO-use}
\begin{figure}[hbt!] 
\centering
\includegraphics[width=1.0\textwidth]{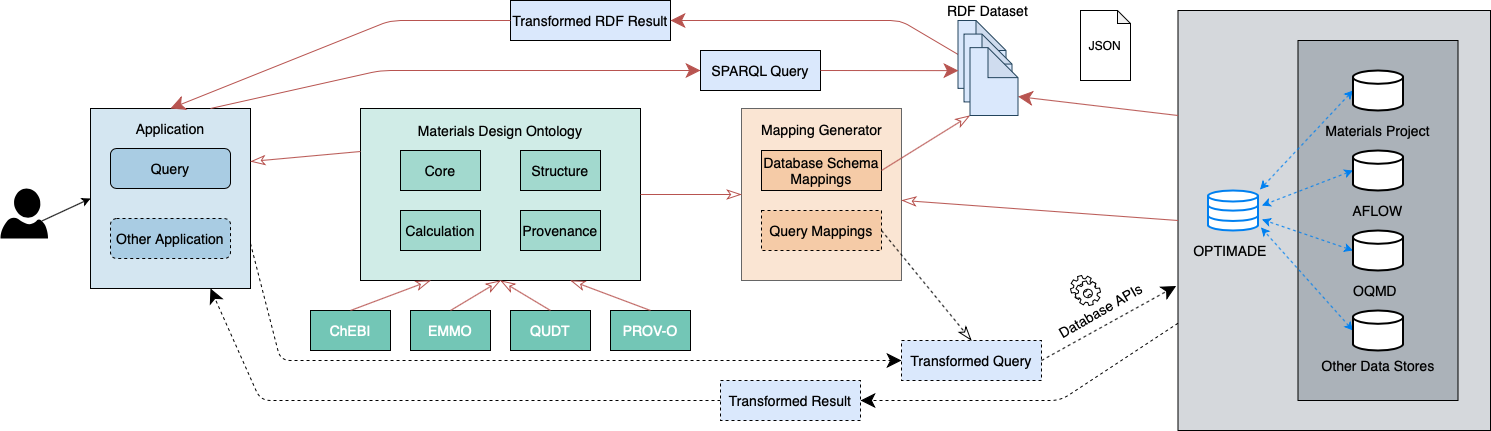}
\caption{The vision of the use of MDO. The full-lined components in the figure are currently implemented in a prototype.}
\label{fig:mdo-vision}
\end{figure}

In Figure \ref{fig:mdo-vision}, we show the vision for the use of MDO for semantic search over OPTIMADE and materials science databases. By generating mappings between MDO and the schemas of materials databases, we can create MDO-enabled query interfaces. The querying can occur, for instance, via MDO-based query expansion, MDO-based mediation or through MDO-enabled data warehouses.

As a proof of concept (full lines in the figure), we created mappings between MDO and the schemas of OPTIMADE and part of Materials Project.
Using the mappings we created an RDF data set with data from Materials project.
Further, we built a SPARQL query application that can be used to query the RDF data set using MDO terminology. Examples are given below.

\noindent
\subsubsection{Instantiating a materials calculation using MDO.}
\begin{figure}[hbt!] 
\centering
\includegraphics[width=1.0\textwidth]{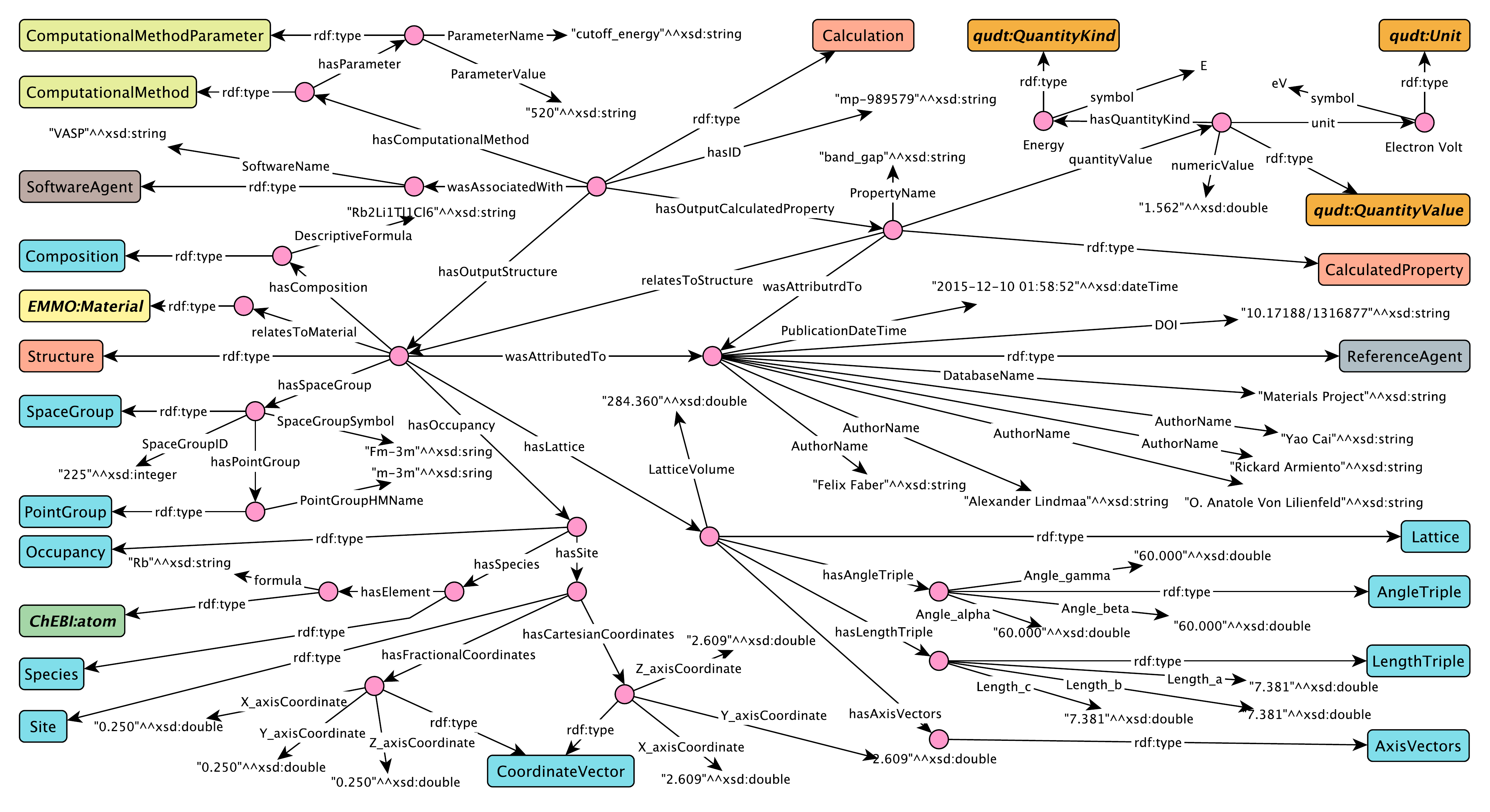}
\caption{An instantiated materials calculation.}
\label{fig:mdo-instance-new}
\end{figure}

In Figure \ref{fig:mdo-instance-new} we exemplify the use of MDO to represent a specific materials calculation and related data in an instantiation. The example is from one of the 85 stable materials published in Materials Project in \cite{faber2016machine}. The calculation is about one kind of elpasolites, with the composition  $\mathrm{Rb}_2\mathrm{Li}_1\mathrm{Ti}_1\mathrm{Cl}_6$. 
To not overcrowd the figure, we only show the instances corresponding to the calculation's output structure, and for multiple calculated properties, species and sites, we only show one instance respectively. 
Connected to the instances of the Core module's concepts, are instances representing the structural information of the output structure, the provenance information of the output structure and calculated property, and the information about the computational method used for the calculation.

\noindent
\subsubsection{Mapping the data from a materials database to RDF using MDO.}
As presented in section \ref{sect-db}, data from many materials databases are provided through the providers' APIs. A commonly used format is JSON.
Our current implementation mapped all JSON data related to the 85 stable materials from \cite{faber2016machine} to RDF. We constructed the mappings by using SPARQL-Generate \cite{lefranccois2017sparql}.
Listing \ref{lst:sg} shows a simple example on how to write the mappings on `band gap' which is a \textit{CalculatedProperty}. The result is shown in Listing \ref{lst:rdf}. The final RDF dataset contains 42,956 triples. The SPARQL-generate script and the RDF dataset are available from the GitHub repository\footnote{\url{https://github.com/huanyu-li/Materials-Design-Ontology/tree/master/mapping_generator}}. This RDF dataset is used for executing SPARQL queries such as the one presented below.

\begin{minipage}[t]{0.49\textwidth}
  \begin{lstlisting}[language=SPARQL, caption=A simple example of mapping, basicstyle=\ttfamily\tiny\color{black}, keywordstyle=\color{eminence}, label = lst:sg]
BASE <@\textcolor{sparqlblue}{<https://w3id.org/mdo/data/1.0/>}@>
PREFIX <@\textcolor{sparqlblue}{fun: <http://w3id.org/sparql-generate/fn/>}@>
PREFIX <@\textcolor{sparqlblue}{core: <https://w3id.org/mdo/core/>}@>
PREFIX <@\textcolor{sparqlblue}{qudt: <http://qudt.org/schema/qudt/}@>
PREFIX <@\textcolor{sparqlblue}{qudt\_unit: <http://qudt.org/vocab/unit/>}@>

<@\textcolor{eminence}{GENERATE}@> {
	?band_gap_node a <@\textcolor{sparqlblue}{core:CalculatedProperty}@>;
	<@\textcolor{sparqlblue}{qudt:quantityValue}@> ?band_gap_quantity_value;
	<@\textcolor{sparqlblue}{core:hasPropertyName}@> <@\textcolor{frenchplum}{"band\_gap"}@>
	<@\textcolor{eminence}{GENERATE}@> {
	    ?band_gap_quantity a <@\textcolor{sparqlblue}{qudt:QuantityValue}@>;
	    <@\textcolor{sparqlblue}{qudt:unit}@> qudt_unit:EV;
	    <@\textcolor{sparqlblue}{qudt:numericValue}@> <@\textcolor{frenchplum}{"band\_gap"}@>
}.
}
<@\textcolor{eminence}{SOURCE}@> <http://example.com/mp-989579_Rb2LiTlCl6.json> 
      <@\textcolor{eminence}{AS}@> ?source
WHERE {
  <@\textcolor{eminence}{BIND}@>(<@\textcolor{sparqlblue}{fun:JSONPath}@>(?source,"$.band_gap") <@\textcolor{eminence}{AS}@> ?band_gap)
  <@\textcolor{eminence}{BIND}@>(<@\textcolor{sparqlblue}{BNODE}@>() <@\textcolor{eminence}{AS}@> ?band_gap_node)
  <@\textcolor{eminence}{BIND}@>(<@\textcolor{sparqlblue}{BNODE}@>() <@\textcolor{eminence}{AS}@> ?band_gap_quantity_value)
}
\end{lstlisting}
  \end{minipage}
  \hfill
  \begin{minipage}[t]{0.49\textwidth}
    \begin{lstlisting}[language=SPARQL, caption=RDF data, basicstyle=\ttfamily\tiny\color{black}, keywordstyle=\color{eminence}, label = lst:rdf]
<@\textcolor{eminence}{@prefix }@> <@\textcolor{sparqlblue}{core: <https://w3id.org/mdo/core/>}@> .
<@\textcolor{eminence}{@prefix }@> <@\textcolor{sparqlblue}{qudt: <http://qudt.org/schema/qudt/}@> .
<@\textcolor{eminence}{@prefix }@> <@\textcolor{sparqlblue}{qudt\_unit: <http://qudt.org/vocab/unit/>}@> .

<@\textcolor{sparqlblue}{<https://w3id.org/mdo/data/1.0/mp-989579\_band\_gap>}@>
        a <@\textcolor{sparqlblue}{core:CalculatedProperty}@> ;
        <@\textcolor{sparqlblue}{core:hasPropertyName}@> <@\textcolor{frenchplum}{"band\_gap"}@> ;
        <@\textcolor{sparqlblue}{qudt:quantityValue}@>[ a <@\textcolor{sparqlblue}{qudt:QuantityValue}@> ;
        <@\textcolor{sparqlblue}{qudt:numericValue}@> <@\textcolor{frenchplum}{1.5623e0}@> ;
        <@\textcolor{sparqlblue}{qudt:unit}@>  <@\textcolor{sparqlblue}{qudt\_unit:EV}@>
        ];
\end{lstlisting}
    \end{minipage}

\noindent
\subsubsection{A SPARQL Query Example.}
As an example, we show a SPARQL query related to CQ6 in Listing \ref{lst:sq}. The result contains 7 records, which are shown in Table \ref{tab:query-result}.
The query is: 
\begin{itemize}
\item ``What are the materials of which the value of band gap is higher than 5eV?" (The result should contain the formula, and the value of band gap.)
\end{itemize}

 \begin{minipage}[c]{0.5\textwidth}
  \begin{lstlisting}[language=SPARQL, caption=A SPARQL query example on Materials Project's dataset, basicstyle=\ttfamily\tiny\color{black}, keywordstyle=\color{eminence},label=lst:sq]
PREFIX <@\textcolor{sparqlblue}{rdf: <http://www.w3.org/1999/02/22-rdf-syntax-ns\#>}@>
PREFIX <@\textcolor{sparqlblue}{core: <https://w3id.org/mdo/core/>}@>
PREFIX <@\textcolor{sparqlblue}{structure: <https://w3id.org/mdo/structure/>}@>
PREFIX <@\textcolor{sparqlblue}{qudt: <http://qudt.org/schema/qudt/>}@>

SELECT ?formula ?value WHERE {
  ?calculation <@\textcolor{sparqlblue}{rdf:type core:Calculation}@>;
   	       <@\textcolor{sparqlblue}{core:hasOutputCalculatedProperty}@> ?property;
   	       <@\textcolor{sparqlblue}{core:hasOutputStructure}@> ?output_structure.
  ?property <@\textcolor{sparqlblue}{qudt:quantityValue}@> ?quantity_value;
   	    <@\textcolor{sparqlblue}{core:hasPropertyName}@> ?name.
  ?quantity_value <@\textcolor{sparqlblue}{rdf:type}@> qudt:QuantityValue;
   	          <@\textcolor{sparqlblue}{qudt:numericValue}@> ?value.
  ?output_structure <@\textcolor{sparqlblue}{structure:hasComposition}@> ?composition.
  ?composition <@\textcolor{sparqlblue}{structure:hasDescriptiveFormula}@> ?formula.
  FILTER (?value><@\textcolor{frenchplum}{5}@> && ?name=<@\textcolor{frenchplum}{"band\_gap"}@>)
} 
\end{lstlisting}
  \end{minipage}
  \hfill
  \begin{minipage}[c]{0.5\textwidth}
    \centering
\scriptsize
\captionof{table}{The result of the query}
\begin{tabular}{l|l} \hline
formula                 & value              \\\hline
$\mathrm{Cs}_2\mathrm{Rb}_1\mathrm{In}_1\mathrm{F}_6$ & 5.3759 \\\hline
$\mathrm{Cs}_2\mathrm{Rb}_1\mathrm{Ga}_1\mathrm{F}_6$  & 5.9392  \\\hline
$\mathrm{Cs}_2\mathrm{K}_1\mathrm{In}_1\mathrm{F}_6$  & 5.4629             \\\hline
$\mathrm{Rb}_2\mathrm{Na}_1\mathrm{In}_1\mathrm{F}_6$  & 5.2687             \\\hline
$\mathrm{Cs}_2\mathrm{Rb}_1\mathrm{Ga}_1\mathrm{F}_6$  & 5.5428             \\\hline
$\mathrm{Rb}_2\mathrm{Na}_1\mathrm{Ga}_1\mathrm{F}_6$  & 5.9026             \\\hline
$\mathrm{Cs}_2\mathrm{K}_1\mathrm{Ga}_1\mathrm{F}_6$   & 6.0426            \\ \hline
\end{tabular}
\label{tab:query-result}
    \end{minipage}
    
We show more SPARQL query examples and the corresponding result in the GitHub repository\footnote{\url{https://github.com/huanyu-li/Materials-Design-Ontology/tree/master/sparql_query}}.

\section{Discussion and Future Work} 
\label{sec-discussion}

To our knowledge, MDO is the first OWL ontology representing solid-state physics concepts, which are the basis for materials design. 
 
The ontology fills a need for semantically enabling access to and integration of materials databases, and for realizing FAIR data in the materials design field.
This will have a large impact on the effectiveness and efficiency of finding relevant materials data and calculations, thereby augmenting the speed and the quality of the materials design process.
Through our connection with OPTIMADE and because of the fact that we have created mappings between MDO and some major materials databases, the potential for impact is large.

The development of MDO followed  well-known practices from the ontology engineering point of view (NeOn methodology and modular design). 
Further, we reused concepts from PROV-O, ChEBI, QUDT and EMMO.
A permanent URL is reserved from w3id.org for MDO.  MDO is maintained on a GitHub repository from where  the ontology in OWL2 DL, visualizations of the ontology and modules, UCs, CQs and restrictions are available. It is licensed via an MIT license\footnote{ \url{https://github.com/huanyu-li/Materials-Design-Ontology/blob/master/LICENSE}}.

Due to our modular approach MDO can be extended with other modules, for instance, regarding different types of calculations and their specific properties.
We identified, for instance, the need for an \textit{X Ray Diffraction} module to model the experimental data of the diffraction used to explore the structural information of materials, and an \textit{Elastic Tensor} module to model data in a calculation that represents a structure's elasticity. 
We may also refine the current ontology. For instance, it may be interesting to model workflows containing \textit{multiple calculations}.

\section{Conclusion}
\label{sec-conclusion}

In this paper, we presented MDO, an ontology which defines concepts and relations to cover the knowledge in the field of materials design and which reuses concepts from other ontologies. We discussed the ontology development process showing use cases and competency questions. 
Further, we showed the use of MDO for semantically enabling materials database search. As a proof of concept, we mapped MDO to OPTIMADE and part of Materials Project and showed querying functionality using SPARQL on a dataset from Materials Project.

\paragraph{{\bf Acknowledgements.}}
This work has been financially supported by the Swedish e-Science Research Centre (SeRC), the Swedish National Graduate School in Computer Science (CUGS), and the Swedish Research Council (Vetenskapsr{\aa}det, dnr 2018-04147).

%
%
%
\bibliographystyle{splncs04}
\bibliography{ref.bib}
\end{document}